\documentclass[amsmath,prb,showpacs]{revtex4}
\usepackage{graphicx}
\usepackage{booktabs}

\begin{document}

\title{Structure and heat capacity of Ne and Xe adsorbed on a bundle of carbon nanotubes}
\author{Daniel E. Shai}
\email{dshai@wooster.edu}
\altaffiliation[Permanent address: ]{College of Wooster, Wooster, OH 44691, USA}
\author{Nathan M. Urban}
\email{nurban@psu.edu}
\altaffiliation[Current address: ]{Department of Geosciences, Pennsylvania State University, University Park, PA 16802, USA}
\author{Milton W. Cole}
\email{miltoncole@aol.com}
\affiliation{Department of Physics, Pennsylvania State University, University Park, Pennsylvania 16802, USA}

\begin{abstract}
The structural and thermal properties of Ne and Xe gases adsorbed
on the outer surface of a large nanotube bundle have been evaluated
with computer simulation. The potential energy model and numerical
techniques were used previously to study Ar [N. M. Urban, S. M.
Gatica, M. W. Cole, and J. L. Riccardo, ``Correlation functions and
thermal properties of Ar adsorbed on the external surface of a
bundle of carbon nanotubes'', Phys. Rev. B \textbf{71}, 245410
(2005)].  Heat capacity results for Ne and Xe exhibit peaks associated
with reordering and ``stripe'' melting transitions for these gases.
\end{abstract}


\maketitle

\section{Introduction}
\label{sec:intro}

In recent years there has been considerable discussion of the
properties of gases adsorbed on bundles of carbon nanotubes.
Measurements of thermal properties (isotherms, isosteric heats, and
thermal desorption behavior) have provided information about the
energies of these gases, while scattering and spectral data have
provided information about their structures and dynamical
properties.\cite{ref1,ref2,ref3,ref4,ref5,ref6,ref7,ref8} In parallel
with these experiments have been various theoretical and simulation
studies, predicting virtually every property that is amenable to
such measurements. Varying assumptions underlying the calculations
pertain to the geometries and the adsorption potentials.

\begin{figure}
\includegraphics[width=8cm]{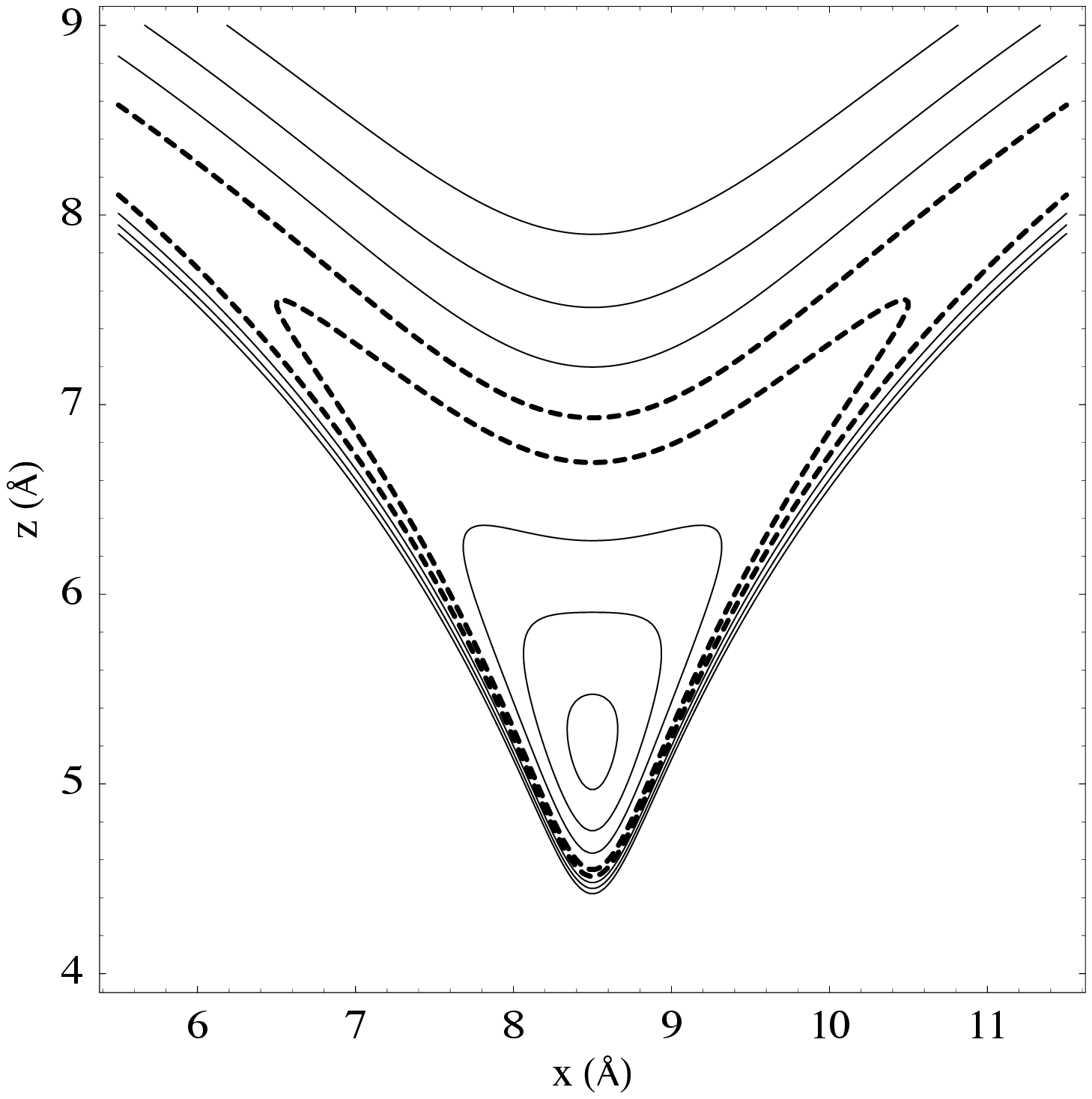}
\includegraphics[width=8cm]{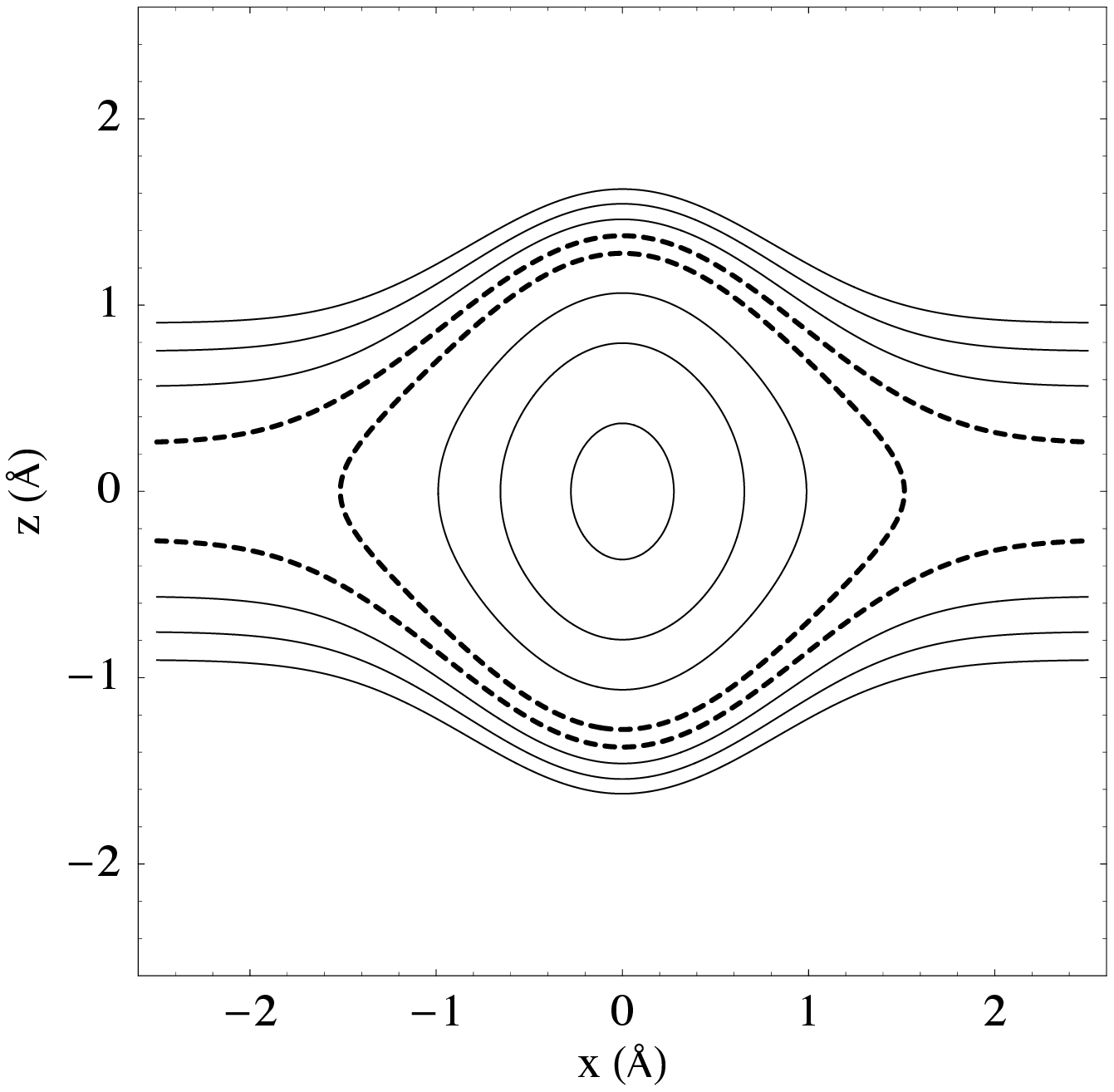}
\caption{\label{fig:1}
Contours of constant potential energy for a Ne atom (a) above the
nanotube surface compared with (b) energy contours generated by the
simple model described in the text. Shown are isopotential lines
for $V=-700$, $-600$, $-500$, $-400$, $-350$, $-300$, $-250$, and
$-200$~K, starting from the groove.  The rapid change in area between
curves for $-400$~K and $-350$~K (dashed lines) manifests a high
density of states, giving rise to a heat capacity peak.}
\end{figure}

Our group has undertaken a series of computer simulation studies
of simple classical gases adsorbed on the outside of large
bundles.\cite{ref9,ref10,ref11}  Since these bundle are large, the
calculations actually assumed the bundle surface to be flat, which
would be precisely correct if the bundle were actually infinite in
radius or else if the bundle consists of a layer of nanotubes
deposited on a surface.  The potential experienced by an adsorbed
molecule was evaluated by summing contributions from a line of
parallel tubes, each of radius $R$; this approach neglects contributions
from the interior of the bundle, an assumption that is plausible
only for the first or second adlayer. Fig.~\ref{fig:1} shows the
potential energy that results for a Ne atom in such an environment,
based on the potential model used in our group. That model ignores
the chirality of the nanotubes by smearing out the carbon atoms
into a cylinder made of continuous carbon matter. These assumptions
are semiquantitatively accurate, at best.

Our simulations employ either canonical or grand canonical Monte
Carlo methods. In early work, the familiar Metropolis algorithm was
used, but our most recent study\cite{ref11} of Ar (denoted I)
employed the efficient Wang-Landau algorithm\cite{ref12}, discussed
below. The adsorption behavior derived from the model consists of
a sequence of ``striped'' phases, with the number of stripes
increasing as the gas pressure $P$ is increased. The first of these,
appearing at low $P$ because of the very attractive potential, is
the so-called ``groove'' phase, in which the gas is confined to a
quasi-one dimensional vicinity of the potential minimum, nestled
between two adjacent tubes. As $P$ increases, a transition to a
``three stripe'' phase occurs, followed by the appearance of a
well defined monolayer. An intriguing ``second layer groove'' phase
has been seen in both simulations and experiments. Multilayer films
can occur, but these have not been investigated because desorption
precludes their observation, except at very low temperature.

This paper focuses on the thermal properties of Ne and Xe, extending
the Ar study, I. The next section summarizes the methodology and
reports results for a very low density phase, called $N=1$. The
number $N$ is the number of adparticles confined within one unit
cell of the simulation (which is periodically replicated across the
surface).  The extent of this cell in the lateral ($x$) direction
is 1.7~nm, taken to be the spacing between (10,10) tubes. In the
$y$ direction (parallel to the tubes' axes), the extent is $10\sigma$,
where $\sigma$ is the Lennard-Jones diameter of the particles; the
other gas parameter entering the calculation is $\epsilon$, the
well depth of the adatom-carbon pair interaction. In the $z$
direction, perpendicular to the bundle surface, the simulation cell
extends a distance $H=4$~nm, where a hard wall terminates the
simulation volume.  Since this boundary underestimates the (macroscopic)
experimental gas volume, its presence tends to suppress desorption
relative to what is seen in experiments.

Sections~\ref{sec:C} and \ref{sec:D} describe simulation results
obtained for Ne and Xe, respectively. Section~\ref{sec:E} summarizes
our principal results and draws conclusions.

\section{Calculations for $N=1$}
\label{sec:B}

\begin{figure}
\includegraphics[width=12cm]{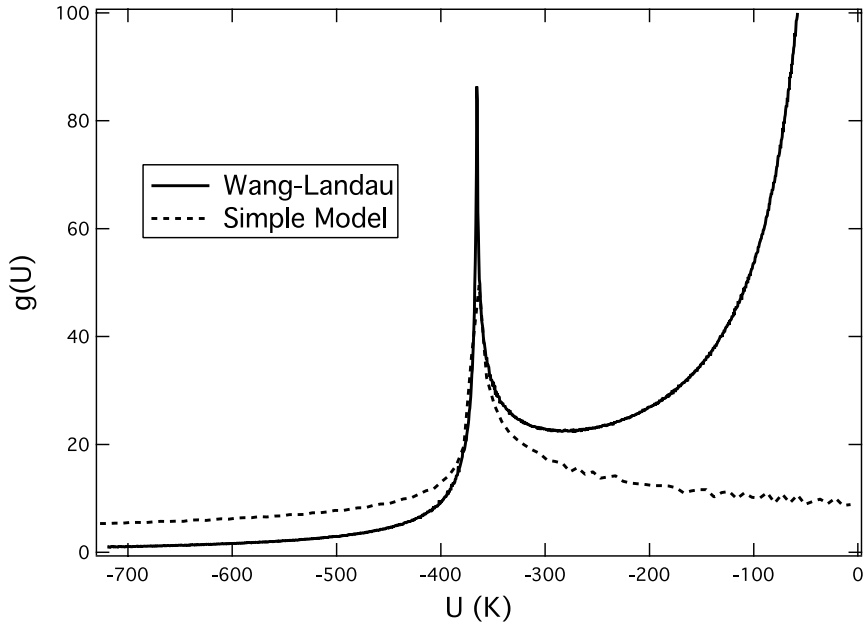}
\caption{\label{fig:2}
Density of states $g(U)$ for $N=1$ Ne atom (full curve). This is
compared with the prediction (dashed curve) derived from the simple
model, Eq.~\ref{eq:4}.}
\end{figure}

The Wang-Landau algorithm is a simulation technique that generates
the density of states $g(U)$, the number of states per unit energy
at energy $U$, for a specified number ($N$) of particles; we omit
a descriptor $N$ from $g(U)$ for notational simplicity. The procedure
used to compute $g(U)$ is discussed in I and in greater detail in
a recent thesis reporting this work.\cite{ref13}  From this function,
thermal properties may be computed at any temperature $T=1/(k_B\beta)$.
Since classical systems have a specified kinetic energy per particle,
$3/(2\beta)$, we focus on the potential energy and omit that
contribution except when showing numerical results for the heat
capacity $C(T)$. For example, the (mean) potential energy $\langle
U(N,T)\rangle$ of the system is
\begin{equation}
\langle U(N,T)\rangle = \int\,U g(U) e^{-\beta U} dU \Big\slash \Bigl[\int\,g(U) e^{-\beta U} dU\Bigr]\,.\label{eq:1}
\end{equation}
The specific heat (in Boltzmann units) is then computed from
\begin{equation}
C/(N k_B) = 3/2 + [\partial U/\partial(N k_B T)]_N\label{eq:2}\,.
\end{equation}
Results for $g(U)$ for $N=1$ are shown in Fig.~\ref{fig:2}.\cite{ref13}
This function exhibits a striking variation with $U$, which can be
understood with simplifying models, as follows. For classical
systems, $g(U)\,dU$ is strictly proportional to the volume of phase
space confined between surfaces of energy $U$ and $U+dU$. This
quantity, for N=1 particle, was called the ``volume density of
states'' in I, where it was evaluated from histograms of the potential
energy distribution in real space; the result agrees with that found
directly with the Wang-Landau simulation method. The lowest energy
``states'' originate from the vicinity of the groove region.  The
resulting low $U$ behavior, $g(U)\sim \mathrm{constant}$, was derived
in I by assuming that the constant energy contours are circles,
centered on the groove; ellipses yield the same analytical form.
The two most striking features of $g(U)$ are a sharp peak, near a
potential energy $U=V_m=-363.373$~K of and a divergence near the
desorption threshold, $U=0$. The latter divergence is of the form
$g(U)\sim |U|^{-4/3}$, where the power 4/3 was derived in I from
the asymptotic van der Waals interaction, $V(z)\sim -1/z^3$. The
divergent behavior just above the peak at $V_m$, $g(U)\sim
(U-V_m)^{-1/2}$ was derived in I, by analyzing an assumed potential
energy near the cylindrical surface at which the monolayer forms:
$V=V_m+(k/2)(z-z_m)^2$.  Here, the value $V_m$ is the holding potential
of a particle near a single cylindrical tube; the quadratic functional
form is analogous to that found near the equilibrium position on a
flat surface. To understand the divergent behavior of $g(U)$ just
below the peak ($U<V_m$), we have invented a simple model potential
that roughly characterizes the spatial dependence near the groove
and monolayer:
\begin{equation}
V_\mathrm{model}(x,z) =  V_m - A \exp{(-x^2)} + (k/2)(z-z_m)^2\,.\label{eq:4}
\end{equation}
Here $A=V_m$ and $k=200$~K/\AA$^2$ are fitting parameters determined
from the potential.  This form is shown in Fig.~\ref{fig:1} to
resemble that of the ``true'' potential near the divergence for
$U\sim V_m$. Even for this simple model, we cannot determine $g(U)$
analytically, unfortunately, but we can evaluate it numerically.
For the ``true'' potential the density of states was calculated
directly using Wang-Landau Monte Carlo, and for the simple model
it was calculated by discretizing space and binning the volume
elements with potentials between $U$ and $U+dU$ to produce a
histogram.  As seen in Fig.~\ref{fig:2}, there is a qualitative
agreement between the two curves for $g(U)$ in the region of the
divergence, just below $U=V_m$.  The divergence originates because
there is a large volume, near the outermost ellipse surrounding the
groove, for which the potential energy is close to this value.

The specific heat of Ne and Xe, for $N=1$, is analogous to that found
for Ar. At low $T$, the value 5/2 is expected for the specific heat
$C/(N k_B)$ because particles have two degrees of freedom due to
potential energy of the confining potential within the groove, plus
three more degrees of freedom from the kinetic energy. The peak in
$C(T)$ near $T=80$~K for Ne, or $T=280$~K for Xe, is due to excitation
out of the groove, on to the cylindrical surface of the nanotube.
This is confirmed by model calculations, as described in I, in which
particles are located either in the groove region or on a nominally
flat surface. The high $T$ limit, $C/(N k_B)=2$, is attributable
to the potential energy contribution from one degree of freedom,
associated with motion perpendicular to the surface.  This behavior
is depicted, along with higher coverage results, in Figs.~\ref{fig:3}
and \ref{fig:4}.

\section{Ne adsorption at finite coverage}
\label{sec:C}

\begin{figure}
\includegraphics[width=12cm]{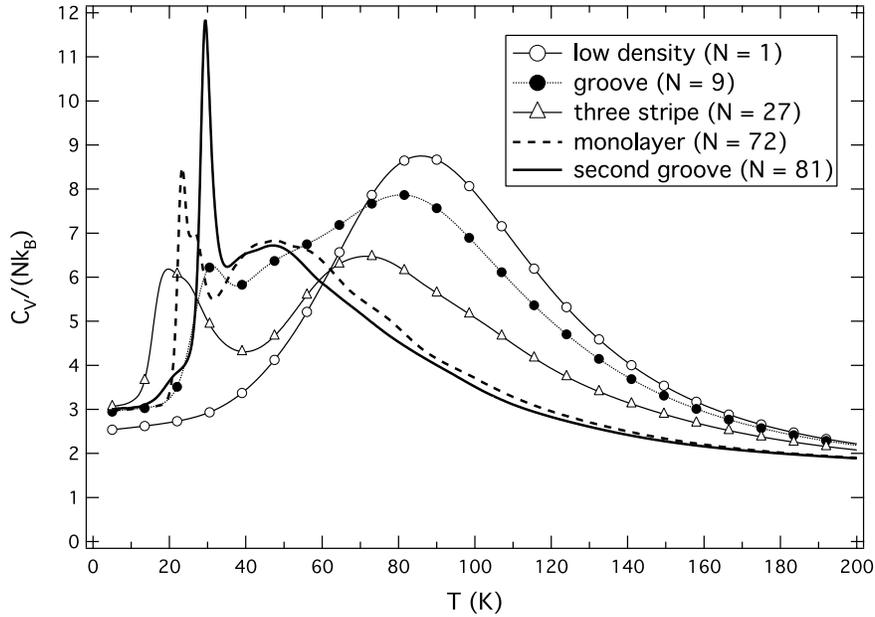}
\caption{\label{fig:3}
Specific heat of Ne for various coverages $N$ (explained in the
text), as a function of $T$.}
\end{figure}

The Wang-Landau method was used to calculate the heat capacity
$C(T)$ for higher coverages (Fig.~\ref{fig:3}).\footnote{The
Wang-Landau algorithm approached convergence for Ne $N=72$, but
ultimately failed to converge completely:  the density of states
at low energies did not achieve sufficient flatness even after the
run time was extended.  Convergence studies suggest that the small
``shoulder'' next to the first heat capacity peak may be a computational
artifact, and the second peak may be slightly larger than is
depicted.} For Ne, the groove phase occurs at a density of $N=9$
adatoms in the simulation cell; the three stripe, monolayer, and
second groove phases occur at $N=27$, 72, and 81, respectively.

The low density $N=1$ results show a heat capacity peak near $T=80$~K,
due to excitation of Ne atoms out of the groove.  The higher coverage
results have a similar peak (referred to here as the second peak),
as well as an additional low temperature peak (first peak), i.e.,
a characteristic double peak structure in $C(T)$.

Monte Carlo configurational samples indicate that the second peak
is due to substantial excitation of adatoms, as in the $N=1$ case.
The excitation is primarily from the groove into the stripe or
monolayer regions.  The second peak occurs at somewhat lower
temperature than in the $N=1$ case, because it is easier to excite
atoms out of a densely packed groove than a nearly empty one. This
is a result of repulsive mutual interactions that predominate when
the groove nears completion.

\begin{figure}
\includegraphics[width=12cm]{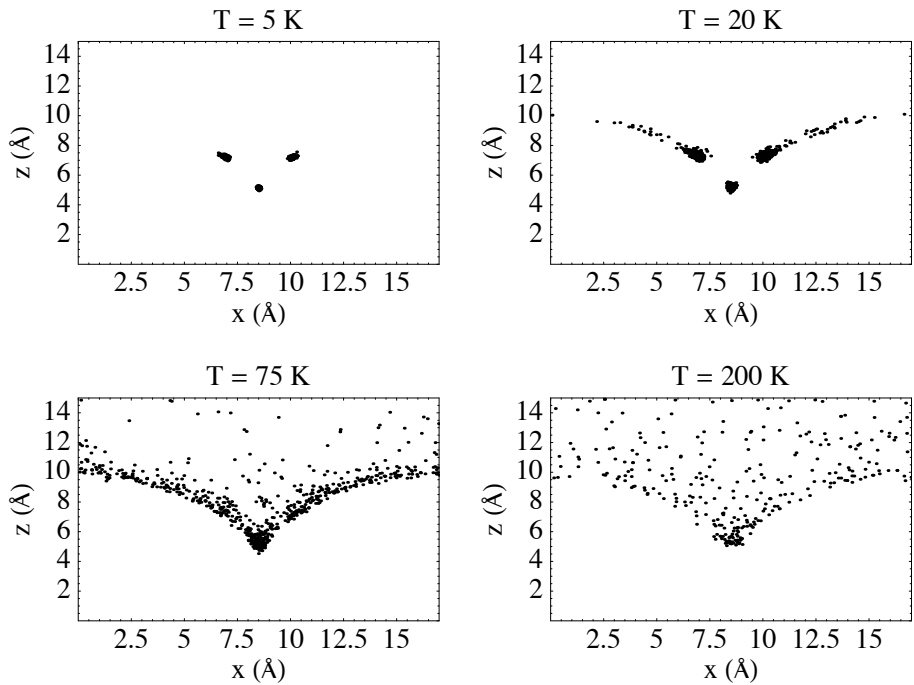}
\caption{\label{fig:4}
Ne density distribution in the three stripe ($N=27$) phase at various
temperatures (cross sectional view of adjacent nanotubes).  The
$T=20$ and 75~K panels show the Ne distribution at the first and
second peaks in the heat capacity.}
\end{figure}

The lower temperature, first peak in the heat capacity, in contrast,
is not associated with promotion of adatoms into different regions
of the substrate.  Rather, it is due to ``energy broadening'', in
which the adatoms remain at or near the same sites, but are thermally
excited over a broader range of energies.  This phenomenon is
exhibited in Fig.~\ref{fig:4} for the three stripe phase.  The
distribution of Ne atoms indicates that at the first peak ($T=20$~K),
most of the atoms remain in the stripes, albeit with some promotion
into the full monolayer.  The atoms remaining in the stripes spread
out, confined less strongly in the stripe potential minima, as a
result of energy broadening.  At the second peak, atoms in the
stripes have been fully promoted into the monolayer and the stripes
disappear.

\section{Xe adsorption at finite coverage}
\label{sec:D}

\begin{figure}
\includegraphics[width=12cm]{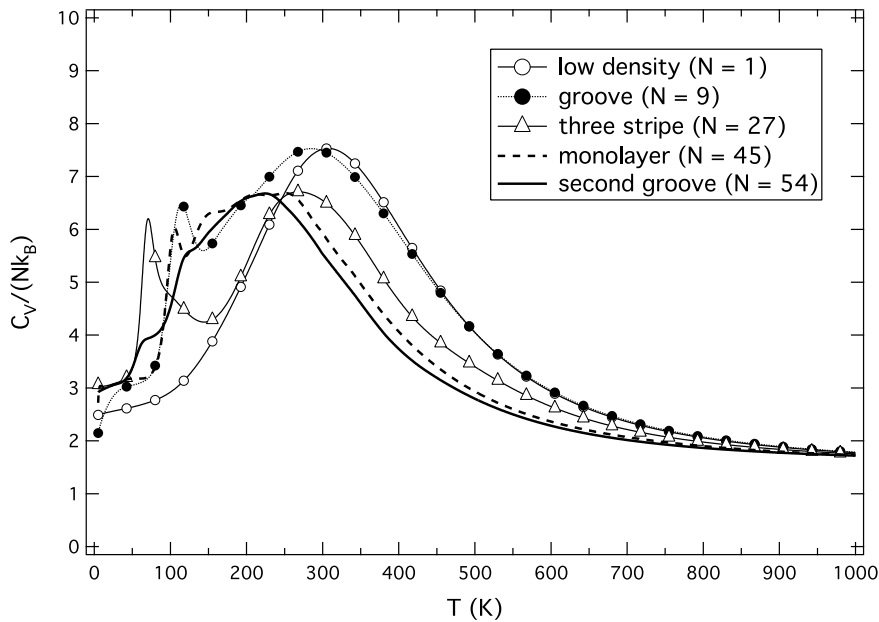}
\caption{\label{fig:5}
Specific heat of Xe for various coverages $N$, as a function of $T$.}
\end{figure}

\begin{figure}
\includegraphics[width=12cm]{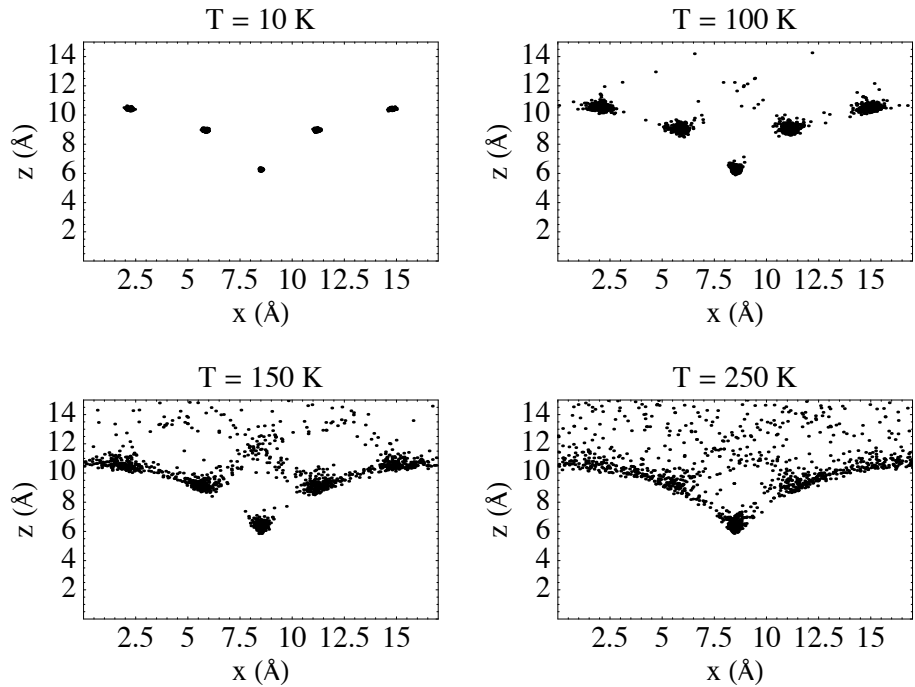}
\caption{\label{fig:6}
Xe density distribution in the monolayer ($N=45$) phase at various
temperatures, shown as cross sectional view of adjacent nanotubes.
The $T=100$ and 250~K panels show the Xe distribution near the first
and second peaks in the heat capacity.}
\end{figure}

The Xe behavior, shown in Figs.~\ref{fig:5} and \ref{fig:6}, is
qualitatively similar to that of Ne, as might be expected. There
are some differences of detail, however. For example, the peaks are
smaller and occur at much higher temperature in the Xe case. The
higher temperature is attributable to the higher energy scale
associated with the interactions' strengths; the gas-gas well depth
for Xe is a factor of seven higher than that of Ne and the adsorption
potential is nearly a factor of three more attractive. (The behavior
of $C(T)$ of Ar, explored in I, is intermediate between these two.)
We surmise that the smaller peak height is due to the fact that the
importance of the adhesive interaction, relative to the cohesive
interaction, is reduced in the Xe case, so the peak in $C(T)$,
attributed primarily to the interaction with the substrate, is less
narrow for Xe. For the same reason, we believe, ``stronger'' cohesive
interactions, arising from subtle intra-film interactions, produce
wiggles for $C(T)$ of Xe, not found for Ne. (We have confirmed that
these are not artifacts of the calculation by refining our simulations.)
We conjecture that the intermediate ``wiggle'' in $C(T)$ for the
monolayer film, near $T=150$~K, is due to the two non-groove stripes
melting into the monolayer (compare the $T=100$ and 150~K density
distributions in Fig.~\ref{fig:6}); the lower-energy groove stripe
itself remains intact until the groove-to-monolayer promotion at
the higher temperature peak near $T=250$~K.

\section{Conclusions}
\label{sec:E}

We have derived the evolution of density with T and the heat capacity
for Ne and Xe gases adsorbed on an array of nanotubes. The behavior
is qualitatively consistent with that reported previously in I for
the case of Ar. These predictions, along with those of the
isotherms,\cite{ref9,ref10,ref11} are testable experimentally and
we look forward to such results, from which conclusions may be
drawn. The principal assumptions that may or may not be validated
by such studies pertain to the geometry and the interactions. The
latter type of assumption would seem more reliable, since the
interactions are semi-empirical, derived from experiments involving
adsorption on graphite. However, extrapolation to the nanotube case,
as done here, represents a further, simplified assumption: that the
chemistry of the atoms is not sensitive to their structure. While
this might seem plausible for C, it remains to be verified. More
severe, presumably, is the assumption that the interaction arises
from a smeared-out array of C atoms. While clearly open to question,
this assumption might be justified by the fact that since the assumed
pair potential is already a rough approximation, the continuum
approximation is not much more drastic. Of course, possible
commensurate structures are not found in the continuum approach.

Concerning the geometry, the assumption of a unique radius for all
tubes is another oversimplification,\cite{ref14} as is the presumption
that the bundle radius is essentially infinite. While more detailed
studies are feasible, they require significantly greater computational
effort.

\begin{acknowledgments}
This research was supported by NSF DMR-0505160 and REU grant
assistance from the Penn State MRSEC program, supported by NSF
DMR-0213623.
\end{acknowledgments}

\end{document}